\documentclass[a4paper,12pt]{article}

\usepackage[left=2.5cm,right=2.5cm,top=2.5cm,bottom=2.5cm]{geometry}
\usepackage{graphicx}
\usepackage{amssymb}
\usepackage{multirow}
\usepackage{amsmath}
\usepackage{psfrag}
\usepackage{setspace}
\usepackage{subcaption}
\usepackage[colorlinks=true,bookmarks=true]{hyperref}
\usepackage{float}
\usepackage{cite}
\usepackage{authblk}
\usepackage{color}

\usepackage{mathtools}

\DeclarePairedDelimiterX\braket[2]{\langle}{\rangle}{#1 \delimsize\vert #2}

\hypersetup{citecolor=blue}
\newcommand{\dif}{\mathrm{d}}
\newcommand{\Eqref}[1]{(\ref{#1})}
\newcommand{\half}{\frac{1}{2}}
\newcommand{\expo}[1]{\mathrm{e}^{#1}}
\newcommand{\brac}[1]{\left(#1 \right)}
\newcommand{\sbrac}[1]{\left[#1\right]}

\newcommand{\Vcal}{\mathcal{V}}
\newcommand{\Lcal}{\mathcal{L}}

\begin{document}

\title{Field equations and particle motion in covariant emergent gravity}

\author[1,2]{Yen-Kheng Lim\footnote{Email: yenkheng.lim@xmu.edu.my}}
\author[1]{Qing-hai Wang\footnote{Email: qhwang@nus.edu.sg}}

\affil[1]{\normalsize{\textit{Department of Physics, National University of Singapore, 117551, Singapore}}}
\affil[2]{\normalsize{\textit{Department of Mathematics, Xiamen University Malaysia, 43900 Sepang, Malaysia}}}

\renewcommand\Authands{ and }

\date{\normalsize{\today}}
\maketitle

\begin{abstract}
  We derive the full set of field equations based on Hossenfelder's recent covariant formulation of the emergent gravity model, along with perturbative and exact solutions. The exact solution describes a static, spherically-symmetric spacetime with a non-trivial vector field which plays the role of dark matter under the emergent gravity paradigm. Equations of motion of relativistic test masses are derived and are shown to reduce to Modified Newtonian Dynamics with additional relativistic corrections. It is also shown that the presence of the vector field gives an additional positive contribution to the bending angle in the deflection of light.
\end{abstract}


\section{Introduction} \label{intro}

Recently, Verlinde \cite{Verlinde:2010hp,Verlinde:2016toy} proposed an interpretation of gravity where it is an emergent process arising out of some underlying microscopic structure. The entropy of the microscopic degrees of freedom appears as the gravitational force in the macroscopic regime. While this idea is fairly new and not free from criticisms \cite{Dai:2017qkz}, it carries many similar features to other approaches attempting to view spacetime as an emergent property arising from (quantum) non-gravitational systems, such as the holographic entanglement entropy \cite{Ryu:2006bv,Hubeny:2007xt,Rangamani:2016dms} entanglement renormalisation  \cite{Vidal:2007hda,Swingle:2009bg}, and exact holographic mapping \cite{Qi:2013caa,Lee:2015vla}. Most of the examples mentioned here are either inspired by or related to the famous AdS/CFT correspondence.

One of the main drawbacks of Verlinde's emergent gravity is that the results are calculated only in the Newtonian limit, and the model does not provide a Lagrangian from which we may derive equations of motion for its variables. This shortcoming has been recently addressed by Hossenfelder who provided a covariant Lagrangian \cite{Hossenfelder:2017eoh} in accordance to Verlinde's model. We shall henceforth refer to this Lagrangian as the Covariant Emergent Gravity (CEG) Lagrangian. 

In this formulation, the theory of emergent gravity is modeled as a typical Einstein-Hilbert action with source terms associated with a vector field $u^\mu$, which was christened the \emph{imposter field} in Ref.~\cite{Hossenfelder:2017eoh}. This imposter field captures the effects of the microscopic degrees of freedom that manifests itself at macroscopic length scales and would tentatively play the role of dark matter and possibly even dark energy. This was further solidified in Ref.~\cite{Hossenfelder:2018vfs} where the Newtonian limit of this CEG Lagrangian is shown to reproduce the acceleration of Modified Newtonian Dynamics (MOND) and is fitted against the galactic rotation curves.

However, the physics arising out of the CEG Lagrangian was also mainly considered in the non-relativistic limit in which the imposter field equations are solved against a flat spacetime, or in the probe limit in which the imposter field is solved on a fixed background metric and back-reactions to the spacetime are ignored. A cosmological spacetime was indeed considered in full relativistic treatment, and a limiting case to the de Sitter solution was obtained. Soon thereafter, Ref.~\cite{Dai:2017guq} pointed out a typo in Eq.~(22) of \cite{Hossenfelder:2017eoh}, and they introduced a small modification to the Lagrangian to obtain another de Sitter limit from a cosmological metric. This CEG Lagrangian should be another relativistic completion to MOND, though so far the derivation of MOND was performed directly in the non-relativistic case of its equations of motion for the imposter field. Therefore, this derived MOND relation would not be able to account for relativistic corrections such as orbital precessions 
and gravitational lensing. 

The aim of this paper is to address these issues with a full relativistic treatment of the equations arising from the CEG Lagrangian. In particular, we find that a full variation of the action without neglecting any terms produces a stress tensor that is different from \cite{Hossenfelder:2017eoh} and \cite{Dai:2017guq}. The difference could be traced to a particular term in the action porportional to  $\delta\Gamma^\lambda_{\mu\nu}$. A full agreement with \cite{Dai:2017guq} is recovered if this term is zero. With this full set of field equations, we
obtain perturbative solutions for spherically-symmetric static spacetimes as well as a time-dependent cosmological spacetime. In the former case of the spherically-symmetric spacetime, we are able to extend the solution to a exact solution to the full equations of motion for a certain choice of Lagrangian parameters. We may interpret this solution as a black hole, at least in the sense that the solution contains a curvature singularity hidden behind a horizon. Indeed, by turning off the imposter field, the spacetime reduces to the Schwarzschild solution.

With the exact solution at hand, we are able to describe the motion of relativistic test masses beyond the Newtonian limit. By Verlinde's and Hossenfelder's construction, the test masses should feel a force coming from the presence of the imposter field. As such, the motion of test particles are no longer described by geodesics of the spacetime, but rather geodesics of an \emph{effective} spacetime where the metric is modified by the imposter field. From its fully relativistic description, we are able to take the Newtonian limit to (re)derive the MOND acceleration along with additional relativistic corrections.

The relativistic solution also allows us to consider gravitational lensing in the spacetime. If we assume that photons are not affected by the imposter field, then the deflection of light could only be caused by the spacetime curvature. Since our solution captures the backreaction effects of the imposter field on spacetime curvature, the deflection of light due to the imposter field occurs only indirectly via this backreaction. As such this model additionally predicts a different amount of lensing compared to standard GR. We shall see below that the strength of the backreaction and the imposter field force on test masses are governed by two independent parameters. Thus one might be hopeful that this additional parameter might accommodate how a typical MOND description is not able to account for strong lensing \cite{Sanders:1998kv}, or, perhaps more generally, how lensing accounts for baryonic matter in the presence of this imposter field \cite{Massey:2010hh}.

The rest of the paper is organised as follows. In Sec.~\ref{Saction} we review the covariant action under emergent gravity and derive its full set of equations of motion. We consider perturbative solutions with a spherically-symmetric and a cosmological ansatz in Sec.~\ref{Psolution}. In Sec.~\ref{Ssolution} we derive an exact solution starting from a general spherically-symmetric ansatz for the spacetime, and by assuming that the imposter field has zero spatial components. Once having the exact solution, we consider the motion of test masses in Sec.~\ref{Stestmass}, and of photons in Sec.~\ref{Sphotons}. We end the paper with some concluding remarks in Sec.~\ref{Sconclusion}.

\section{Action and equations of motion}\label{Saction}

Let us briefly review the essential features of Hossenfelder's CEG action. The basic variables are the metric $g_{\mu\nu}$ and the imposter field $u^\mu$. 
By dimensional analysis, Hossenfelder argues that the Lagrangian should be of the form $\chi^{3/2}$, where $\chi$ is a term that is quadratic in derivatives of $u$, namely, $\chi\sim\brac{\nabla u}^2$. Therefore, the Lagrangian for the imposter field should take the form
\begin{align}
 \mathcal{L}_{\theta}=\frac{\alpha}{16\pi G}\chi^{3/2}-\Vcal(u),
\end{align}
where $\Vcal(u)$ is a potential described as a function of $u=\sqrt{-u_\mu u^\mu}$ and was chosen differently in previous literature. In particular, Hossenfelder chose $\Vcal(u)\propto u^2$ in \cite{Hossenfelder:2017eoh}, while Dai and Stojkovic had $\Vcal(u)\propto u^4$ in \cite{Dai:2017guq}. The parameter $\alpha$ characterises the coupling strength of the imposter field to gravity, and is expected to be of an order of inverse cosmological length scales. At this stage, the term $\chi$ is only required to be quadratic in $\nabla u$, which leads to three possible contractions
\begin{align}
 \chi&=\bar{a}\nabla_\sigma u^\sigma\nabla_\lambda u^\lambda+\bar{b}\nabla_\sigma u_\lambda\nabla^\sigma u^\lambda+\bar{c}\nabla_\sigma u_\lambda\nabla^\lambda u^\sigma.\label{def_chi}
\end{align}
In Refs.~\cite{Hossenfelder:2017eoh} and \cite{Dai:2017guq}, a specific choice was made for the coefficients of the kinetic terms, namely $\bar{a}=\frac{4}{3},\,\bar{b}=\bar{c}=-\half$.
However, in the following, we shall regard $\bar{a}$, $\bar{b}$, and $\bar{c}$ as arbitrary coefficients which may take other possible values. 

Next, we wish to determine the Lagrangian to describe the interaction between normal matter and the imposter field. The main idea of Verlinde and Hossenfelder is that the effects normally attributed to dark matter is due to forces arising from the interaction between normal matter and the imposter field. In other words, the imposter field couples with the stress tensor of normal matter defined by
\begin{align}
 T_{\mu\nu}=\Lcal_{\mathrm{m}} g_{\mu\nu}-2\frac{\delta\Lcal_{\mathrm{m}}}{\delta g^{\mu\nu}}, \label{Tmunu}
\end{align}
where $\Lcal_{\mathrm{m}}$ is the Lagrangian of normal matter. To construct the specific form of the interaction Lagrangian/action, we revisit the main idea of Verlinde and Hossenfelder, in which normal matter feels an effective metric of the form
\begin{align}
 \tilde{g}_{\mu\nu}=g_{\mu\nu}-\beta\frac{u_\mu u_\nu}{u}. \label{tildeg}
\end{align}
On this basis, we can construct the interaction Lagrangian by considering the motion of a time-like test particle of mass $m$ (made with normal matter) with trajectory $x^\mu(\tau)$, where $\tau$ parameterises the trajectory. The action and corresponding stress tensor for the particle are\footnote{See, for instance, \cite{hobson2006general}.}
\begin{align}
 I_{\mathrm{m}}&=\frac{m}{2}\int\dif\tau g^{\mu\nu}\dot{x}_\mu\dot{x}_\nu,\quad T_{\mu\nu}(y)=-\frac{m}{\sqrt{-g(y)}}\int\dif\tau\,\delta(x-y)\dot{x}_\mu\dot{x}_\nu,
\end{align}
where over-dots denote derivatives with respect to $\tau$. If the test particle (made from normal matter) is to feel an effective metric \Eqref{tildeg}, our desired $(\mathrm{matter})+(\mathrm{interaction})$ should be
\begin{align}
 I_{\mathrm{m}}+I_{\mathrm{int}}=\frac{m}{2}\int\dif\tau\,\tilde{g}_{\mu\nu}\dot{x}^\mu\dot{x}^\nu=\frac{m}{2}\int\dif\tau\,\brac{g_{\mu\nu}-\beta\frac{u_\mu u_\nu}{u}}\dot{x}^\mu\dot{x}^\nu. \label{TestParticleAction}
\end{align}
This would be achieved if the interaction takes the form
\begin{align}
 I_{\mathrm{int}}=\frac{\beta}{2}\int\dif^4x\sqrt{-g}\frac{u^\mu u^\nu}{u}T_{\mu\nu}=\int\dif^4x\sqrt{-g}\,\Lcal_{\mathrm{int}},
\end{align}
where
\begin{align}
 \mathcal{L}_{\mathrm{int}}=\frac{\beta}{2}\frac{u^\mu u^\nu}{u}T_{\mu\nu}.
\end{align}

Assembling the pieces together, the CEG model is described by the action
\begin{align}
 I&=\int\dif^4x\sqrt{-g}\;\brac{\frac{1}{16\pi G}R+\mathcal{L}_{\mathrm{m}}+\mathcal{L}_\theta+\mathcal{L}_{\mathrm{int}}}. \label{action}
\end{align}
Here, we note that in our derivation of $\Lcal_{\mathrm{int}}$, we have introduced a small modification to Hossenfelder's action. Namely, there is an additional factor of $\half$ when comparing Eq.~\Eqref{action} to Eq.~(6) of \cite{Hossenfelder:2017eoh}. (In the present notation, $\beta=1/L$, where $L$ is the notation used by Hossenfelder.) However, we argue that this factor is necessary for Eq.~\Eqref{TestParticleAction}, or equivalently, Eq.~(5) of \cite{Hossenfelder:2017eoh}, to hold.

In the following, we find it easier to keep track of the terms by manipulating the symmetric and anti-symmetric parts of $\chi$ separately. As such we follow \cite{Hossenfelder:2017eoh} and consider the \emph{strain tensor} defined by 
\begin{align}
 \epsilon_{\mu\nu}=\nabla_\mu u_\nu+\nabla_\nu u_\mu,
\end{align}
in addition to an anti-symmetric combination
\begin{align}
 F_{\mu\nu}=\nabla_\mu u_\nu-\nabla_\nu u_\mu.
\end{align}
We shall also redefine our coefficients in Eq.~\Eqref{def_chi} by 
\begin{align}
 a=\frac{\bar{a}}{2},\quad b=\frac{\bar{b}+\bar{c}}{2},\quad c=\frac{\bar{b}-\bar{c}}{2}. \label{def_abc}
\end{align}
In terms of these quantities, Eq.~\Eqref{def_chi} becomes
\begin{align}
 \chi=\frac{a}{2}\brac{{\epsilon^\sigma}_\sigma}^2+\frac{b}{2}\epsilon^{\sigma\lambda}\epsilon_{\sigma\lambda}+\frac{c}{2}F^{\sigma\lambda}F_{\sigma\lambda}.
\end{align}

In performing the variation of the action, a crucial ingredient involves the variation of $\chi$, which is given by
\begin{align}
 \delta\chi&=A_{\mu\nu}\delta g^{\mu\nu}+2B^{\mu\nu}\brac{\nabla_\mu \delta u_\nu-u_\lambda\delta\Gamma^{\lambda}_{\mu\nu}},\label{deltachi}
\end{align}
where 
\begin{subequations}
\begin{align}
 A_{\mu\nu}&=a{\epsilon^\lambda}_\lambda\epsilon_{\mu\nu}+b\epsilon_{\mu\lambda}{\epsilon_\nu}^\lambda+cF_{\mu\lambda}{F_\nu}^\lambda,\\
 B^{\mu\nu}&=a{\epsilon^\lambda}_\lambda g^{\mu\nu}+b\epsilon^{\mu\nu}+cF^{\mu\nu}.
\end{align}
\end{subequations}
Hence, the variation of the action is 
\begin{align}
 \delta I&=\frac{1}{16\pi G}\int\dif^4x\sqrt{-g}\Biggl\{3\alpha\chi^{1/2}B^{\mu\nu}\nabla_\mu\delta u_\nu+16\pi G\frac{\dif\Vcal}{\dif u}\frac{u^\nu}{u}\delta u_\nu\nonumber\\
         &\hspace{1.5cm}+8\pi G\beta\brac{\frac{u^\sigma u^\lambda T_{\sigma\lambda}u^\nu}{u^3}+\frac{2u^\lambda {T_{\lambda}}^\nu}{u}}\delta u_\nu\nonumber\\
         &\hspace{1.5cm}+\brac{R_{\mu\nu}-\half Rg_{\mu\nu}-8\pi G T_{\mu\nu}}\delta g^{\mu\nu}+g_{\mu\nu}\nabla_\sigma\nabla^\sigma\delta g^{\mu\nu}-\nabla_\mu\nabla_\nu\delta g^{\mu\nu}\nonumber\\
         &\hspace{1.5cm}+8\pi G\beta\brac{\frac{u^\lambda u^\sigma T_{\lambda\sigma}u_\mu u_\nu}{2u^3}-\frac{u^\lambda u^\sigma T_{\lambda\sigma}}{2u}g_{\mu\nu}+\frac{2u_\mu u^\lambda T_{\lambda\nu}}{u}}\delta g^{\mu\nu}\nonumber\\
         &\hspace{1.5cm}+\sbrac{\alpha\chi^{1/2}\brac{\frac{3}{2}A_{\mu\nu}-\half\chi g_{\mu\nu}}+8\pi G\Vcal g_{\mu\nu}+8\pi G\frac{\dif\Vcal}{\dif u}\frac{u_\mu u_\nu}{u}}\delta g^{\mu\nu}\nonumber\\
         &\hspace{1.5cm}+\frac{3\alpha}{2}u_\mu\chi^{1/2}{B^\sigma}_\nu\nabla_\sigma\delta g^{\mu\nu}+\frac{3\alpha}{2}u_\nu\chi^{1/2}{B_\mu}^\sigma\nabla_\sigma\delta g^{\mu\nu}-\frac{3\alpha}{2}u^\lambda\chi^{1/2}B_{\mu\nu}\nabla_\lambda\delta g^{\mu\nu}\Biggr\}.\label{deltaL}
\end{align}
Note that we did not vary $T_{\mu\nu}$ itself in $\Lcal_{\mathrm{int}}$.

Before proceeding, let us take a moment to draw a comparison between Eq.~\Eqref{deltaL} and the results of \cite{Hossenfelder:2017eoh,Dai:2017guq}, particularly the terms involving the variation $\delta g^{\mu\nu}$. Clearly the third line in Eq.~\Eqref{deltaL} is the Einstein tensor and the stress tensor due to normal matter (plus its corresponding boundary term). The fourth line is precisely $(T_{\mathrm{int}})_{\mu\nu}$ as given by \cite{Dai:2017guq}. The fifth line has the term
\begin{align}
 -\alpha\chi^{1/2}\brac{\frac{3}{2}A_{\mu\nu}-\half\chi g_{\mu\nu}}&=-\alpha\chi^{1/2}\brac{\frac{3a}{2}{\epsilon^\lambda}_\lambda\epsilon_{\mu\nu}+\frac{3b}{2}\epsilon_{\mu\lambda}{\epsilon_\nu}^\lambda+\frac{3c}{2}F_{\mu\lambda}{F_\nu}^\lambda-\half\chi g_{\mu\nu}}.
\end{align}
Now, Hossenfelder's choice of parameters were $\bar{a}=\frac{4}{3}$, $\bar{b}=\bar{c}=-\frac{1}{2}$. Via Eq.~\Eqref{def_abc}, this corresponds to $a=\frac{2}{3}$, $b=-\frac{1}{2}$, and $c=0$. With this choice, the above equation becomes
\begin{align}
 \frac{\alpha}{2}\chi^{1/2}\brac{-2{\epsilon^\lambda}_\lambda\epsilon_{\mu\nu}+\frac{3}{2}\epsilon_{\mu\lambda}{\epsilon_\nu}^\lambda+\chi g_{\mu\nu}}.
\end{align}
This term, when added to $8\pi G\brac{\Vcal g_{\mu\nu}+\frac{\dif\Vcal}{\dif u}\frac{u_\mu u_\nu}{u}}$, is precisely $\brac{T_s}_{\mu\nu}$ as given by \cite{Dai:2017guq}. We have already reproduced all the stress tensor components of \cite{Dai:2017guq}, but the last line of Eq.~\Eqref{deltaL} is still unaccounted for!

To see where this line came from, we recall that the last term of Eq.~\Eqref{deltachi} involves the variation
\begin{align}
 \delta\Gamma^\lambda_{\mu\nu}&=-\half\brac{g_{\mu\sigma}\nabla_\nu\delta g^{\sigma\lambda}+g_{\nu\sigma}\nabla_\mu\delta g^{\sigma\lambda}-g_{\mu\sigma}g_{\nu\rho}\nabla^\lambda\delta g^{\sigma\rho}}.
\end{align}
This term contributes to the last line of Eq.~\Eqref{deltaL}. If $\delta\Gamma^\lambda_{\mu\nu}=0$, or if the last line of Eq.~\Eqref{deltaL} is not present, the variation $\frac{\delta I}{\delta g^{\mu\nu}}=0$ reproduces exactly the stress tensor given in \cite{Dai:2017guq}. However, we are unable to find any justification to neglect these terms. Even if we were to specialise to Hossenfelder and Dai et al's choice of $a=\frac{2}{3}$, $b=-\frac{1}{2}$, and $c=0$, this term is still non-zero.

Keeping all the terms in Eq.~\Eqref{deltaL} and performing integration by parts, the variation of the action is
\begin{align}
 \delta I&=\frac{1}{16\pi G}\int\dif^4x\sqrt{-g}\Biggl\{-\delta u_\nu 3\alpha\nabla_\mu\brac{\chi^{1/2}B^{\mu\nu}}+16\pi G\frac{\dif\Vcal}{\dif u}\frac{u^\nu}{u}\delta u_\nu\nonumber\\
  &\quad+8\pi G\beta\brac{\frac{u^\sigma u^\lambda T_{\sigma\lambda}u^\nu}{u^3}+\frac{2u^\lambda {T_{\lambda}}^\nu}{u}}\delta u_\nu+\brac{R_{\mu\nu}-\half Rg_{\mu\nu}-8\pi GT_{\mu\nu}}\delta g^{\mu\nu}\nonumber\\
     &\quad+8\pi G\beta\brac{\frac{u^\lambda u^\sigma T_{\lambda\sigma}u_\mu u_\nu}{2u^3}-\frac{u^\lambda u^\sigma T_{\lambda\sigma}}{2u}g_{\mu\nu}+\frac{2u_\mu u^\lambda T_{\lambda\nu}}{u}}\delta g^{\mu\nu}\nonumber\\
     &\quad+\sbrac{\alpha\chi^{1/2}\brac{\frac{3}{2}A_{\mu\nu}-\half\chi g_{\mu\nu}}+8\pi G\Vcal g_{\mu\nu}+8\pi G\frac{\dif\Vcal}{\dif u}\frac{u_\mu u_\nu}{u} }\delta g^{\mu\nu}\nonumber\\
     &\quad-\frac{3\alpha}{2}\sbrac{\nabla_\sigma\brac{u_\mu\chi^{1/2}{B^\sigma}_\nu}+\nabla_\sigma\brac{u_\nu\chi^{1/2}{B_\mu}^\sigma}-\nabla_\lambda\brac{u^\lambda\chi^{1/2}B_{\mu\nu}}}\delta g^{\mu\nu}\nonumber\\
     &\quad+\nabla_\sigma\brac{g_{\mu\nu}\nabla^\sigma\delta g^{\mu\nu}-\nabla_\nu\delta g^{\sigma\nu}}\nonumber\\
     &\quad+3\alpha\nabla_\sigma\sbrac{\chi^{1/2}\brac{B^{\sigma\nu}\delta u_\nu+\half\brac{u_\mu {B^\sigma}_\nu+u_\nu {B_\mu}^\sigma-u^{\sigma}B_{\mu\nu}}\delta g^{\mu\nu}}}\Biggr\},
\end{align}
where the last two lines are total derivatives which only provide contributions to the boundary of the spacetime. The third-last line is proportional to $\delta g^{\mu\nu}$, and would contribute to the stress tensor, ultimately modifying the stress tensor used by \cite{Dai:2017guq} and \cite{Hossenfelder:2017eoh}.

The variation $\frac{\delta I}{\delta u^\nu}=0$ gives the equation of motion for $u^\mu$, which we shall refer to as the \emph{imposter equation},
\begin{align}
 \frac{3\alpha}{16\pi G}\nabla_\mu\sbrac{\chi^{1/2}\brac{a{\epsilon^\sigma}_\sigma g^{\mu\nu}+b\epsilon^{\mu\nu}+cF^{\mu\nu}}}=\frac{\dif\Vcal}{\dif u}\frac{u^\nu}{u}+\frac{\beta}{2}\brac{\frac{2u^\lambda {T_\lambda}^\nu}{u}+\frac{u^\sigma u^\lambda T_{\sigma\lambda} u^\nu}{u^3}}.\label{EOMu}
\end{align}
Finally, the variation $\frac{\delta I}{\delta g^{\mu\nu}}=0$ gives us the Einstein equation
\begin{align}
 R_{\mu\nu}-\half Rg_{\mu\nu}&=8\pi G\brac{T_{\mu\nu}+\brac{T_{\mathrm{int}}}_{\mu\nu}+\brac{T_s}_{\mu\nu}}\nonumber\\
    &\quad-\frac{3\alpha}{2}\sbrac{\nabla_\sigma\brac{u_\mu\chi^{1/2}{B^\sigma}_\nu}+\nabla_\sigma\brac{u_\nu\chi^{1/2}{B_\mu}^\sigma}-\nabla_\lambda\brac{u^\lambda\chi^{1/2}B_{\mu\nu}}},\label{EE1}
\end{align}
where 
\begin{align}
 \brac{T_{\mathrm{int}}}_{\mu\nu}&=\beta\brac{\frac{u^\lambda u^\sigma T_{\lambda\sigma}u_\mu u_\nu}{2u^3}-\frac{u^\lambda u^\sigma T_{\lambda\sigma}}{2u}g_{\mu\nu}+\frac{2u_\mu u^\lambda T_{\lambda\nu}}{u}},\\
 \brac{T_s}_{\mu\nu}&=\sbrac{\frac{\alpha}{8\pi G}\chi^{1/2}\brac{\frac{3}{2}A_{\mu\nu}-\half\chi g_{\mu\nu}}+\Vcal g_{\mu\nu}+\frac{\dif\Vcal}{\dif u}\frac{u_\mu u_\nu}{u} }.
\end{align}
With the resulting Einstein equations, we can now reiterate our comparison with Dai and Stojkovic's stress tensor \cite{Dai:2017guq} more explicitly. Putting in Hossenfelder's initial choice $\bar{a}=\frac{4}{3}$, $\bar{b}=\bar{c}=-\half$ (which translates to $a=\frac{2}{3}$, $b=-\half$, and $c=0$) along with a quartic potential for $\Vcal$, the $(T_{\mu\nu}+\brac{T_{\mathrm{int}}}_{\mu\nu}+\brac{T_s}_{\mu\nu})$ of Eq.~\Eqref{EE1} agrees with  those of \cite{Dai:2017guq} exactly. But we have additional terms on the second line of \Eqref{EE1} which was previously unaccounted for. This originated from the $\delta\Gamma$ variation which we are unable to justify neglecting.

We can rewrite Eq.~\Eqref{EE1} in a different form by using Eq.~\Eqref{EOMu} to simplify some terms involving the potential and stress tensor\footnote{Our notation for symmetrisation is $A_{(\mu\nu)}=\half\brac{A_{\mu\nu}+A_{\nu\mu}}$.}
\begin{align}
 &R_{\mu\nu}-\half Rg_{\mu\nu}=8\pi G\sbrac{T_{\mu\nu}+\half\beta u^\lambda u^\sigma T_{\lambda\sigma}\brac{g_{\mu\nu}+\frac{u_\mu u_\nu}{u^2}}+\frac{\dif\Vcal}{\dif u}\frac{u_\mu u_\nu}{u}-\Vcal g_{\mu\nu}}\nonumber\\
 &\hspace{3cm}-\alpha c\chi^{1/2}\brac{\frac{3}{2}F_{\mu\lambda}{F_\nu}^\lambda-\frac{1}{4}F^{\sigma\lambda}F_{\sigma\lambda}g_{\mu\nu}}\nonumber\\
 &\hspace{3cm}+\alpha\chi^{1/2}\brac{-\frac{a}{2}\brac{{\epsilon^\sigma}_\sigma}^2g_{\mu\nu}+\frac{b}{4}\epsilon^{\sigma\lambda}\epsilon_{\sigma\lambda}g_{\mu\nu}-\frac{3b}{4}{\epsilon^\lambda}_\lambda\epsilon_{\mu\nu}+\frac{3b}{2}F_{\sigma(\mu}{\epsilon^\sigma}_{\nu)}}\nonumber\\
 &\hspace{3cm}-\frac{3\alpha}{2}u^\lambda\nabla_\lambda\sbrac{\chi^{1/2}\brac{a{\epsilon^\sigma}_\sigma g_{\mu\nu}+b\epsilon_{\mu\nu}}}-3\alpha c\sbrac{\nabla_\sigma\brac{\chi^{1/2}{F^\sigma}_{(\mu}}}u_{\nu)}.\label{EOMg}
\end{align}

\section{Perturbative solutions}
\label{Psolution}
In this section, we try to solve the equations of motion in Eqs.~(\ref{EOMu}) and (\ref{EOMg}) perturbatively. We assume that the metric is not far from the flat spacetime and the impostor field strength is weak. We only consider two types of metrics. One is static spherical-symmetric, another one is the Robertson-Walker metric. For the impostor field, we assume that the temporal component dominant and all spatial components are negligible.  

For simplicity, we consider the absence of normal matter ($T_{\mu\nu}=0$). We assume the imposter field potential $\Vcal$ to take the most general quartic form, 
\begin{equation}
\Vcal(u)= \lambda'_0 + \lambda'_1 u + \lambda'_2 u^2 + \lambda'_3 u^3 + \lambda'_4 u^4.
\end{equation}
Note that we can always shift $u$ to eliminate the cubic term. That is 
\begin{equation}
\Vcal(u)= \lambda_0 + \lambda_1 u + \lambda_2 u^2 + \lambda_4 u^4.
\end{equation}
In both types of metric considered in this section, we found that $\Vcal$ must be pure quartic. That is, $\lambda_0=\lambda_1=\lambda_2=0$. This is in agreement with Ref.~\cite{Dai:2017guq}. Thus, we will only consider 
\begin{equation}
\Vcal(u)= \lambda u^4
\end{equation}
in the rest of the paper.

\subsection{Spherically-symmetric solution} \label{SSssph}
Consider a spherically-symmetric and static ansatz of the form
\begin{subequations}
	\begin{align}
	\dif s^2&=-f(r)\dif t^2+\frac{\dif r^2}{h(r)}+r^2\dif\Omega^2_{(2)},\label{metric_ansatz}\\
	u^\mu&=\phi(r)\delta^\mu_t, \label{ansatz}
	\end{align}
\end{subequations}
where $f$, $h$, and $\phi$ are scalar functions that depend only on $r$. Before presenting an exact solution for a particular choice of parameters of $a$, $b$, and $c$ in the next section, let us discuss perturbative solutions for generic parameters. Consider perturbative expansions of the form
\begin{subequations}
	\begin{align}
	f(r) &\sim 1 + f_1(r) \delta+ \cdots, &\delta\to0,\label{f_pert}\\
	h(r) &\sim 1 + h_1(r) \delta + \cdots, &\delta\to0,\label{h_pert}\\
	\phi(r) &\sim \phi_1(r) \epsilon  + \cdots, &\epsilon\to0,\label{phi_pert}
	\end{align}
\end{subequations}
where $\delta$ and $\epsilon$  are two small dimensionless parameters. Different choices of relative sizes of $\delta$ and $\epsilon$ lead to different perturbative solutions. For example, suppose that $\delta$ and $\epsilon$ are in the same order, without loss of generality, consider $\delta=\epsilon$. We get
\begin{subequations}
	\begin{align}
	f(r) &\sim 1 -\left(C_1 + \frac{2M}{r}\right) \epsilon + \frac{2C_1M}{r}\epsilon^2 - \alpha q^3 \sqrt{-(b+c)} \left[\frac{3b}{r} + (2b-c) \frac{\ln r}{r}\right] \epsilon^3 + \mathcal{O}(\epsilon^4), \label{f_pert1}\\
	h(r) &\sim 1 - \frac{2M}{r} \epsilon - \alpha q^3 (2b-c) \sqrt{-(b+c)}  \left(\frac{\ln r}{r}\right) \epsilon^3  + \mathcal{O}(\epsilon^4), 
	\label{h_pert1}\\
	\phi(r) & \sim q \left(\ln\frac{r}{r_0}\right) \epsilon + \left\{\frac{Mq}{(b+c)} \left(-\frac{3b}{r} + \frac{2c}{r} \ln\frac{r}{r_0}\right)\right. \nonumber\\
	&\qquad \left. + \frac{2\lambda q^2 r^3}{243\alpha[-(b+c)]^{3/2}} \left[8 -18 \ln\frac{r}{r_0} +18 \left(\ln\frac{r}{r_0}\right)^2 -9 \left(\ln\frac{r}{r_0}\right)^3\right] \right\}\epsilon^2 
+ \mathcal{O}(\epsilon^3),\label{phi_pert1}
	\end{align}
\end{subequations}
where $M$, $C_1$, $r_0$, and $q$ are integration constants. Obviously that $b+c<0$ is required for the solutions to be real. We found that this is a generic constraint for the parameters for any static and spherical-symmetric metric. Meanwhile, the parameter $a$ disappears from this type of metric. Note that the impostor field affects the metric starting from the order of $\epsilon^3$. In this sense, the $\mathcal{O}(\epsilon)$ and $\mathcal{O}(\epsilon^2)$ terms in the metric can be regarded as background gravitation, something is similar to the Schwarzschild solution, which is independent of the impostor field.  

Another choice would be the one similar to Ref.\cite{Hossenfelder:2017eoh}, suppose that $\delta =\epsilon^2$. In this case, we get
\begin{subequations}
	\begin{align}
	f(r) &\sim 1 -\left(C_1 + \frac{2M}{r}\right) \epsilon^2 - \alpha q^3 \sqrt{-(b+c)} \left[-3b + \frac{3b}{r} + (2b-c) \frac{\ln r}{r}\right] \epsilon^3
+ \mathcal{O}(\epsilon^4), \label{f_pert2}\\
	h(r) &\sim 1 - \frac{2M}{r} \epsilon^2 - \alpha q^3 (2b-c) \sqrt{-(b+c)}  \left(\frac{\ln r}{r}\right) \epsilon^3  + \mathcal{O}(\epsilon^4), 
	\label{h_pert2}\\
	\phi(r) & \sim q \left(\ln\frac{r}{r_0}\right) \epsilon  + \frac{2\lambda q^5r^3 }{243\alpha[-(b+c)]^{3/2}} \left[8 -18 \ln\frac{r}{r_0} +18 \left(\ln\frac{r}{r_0}\right)^2 -9 \left(\ln\frac{r}{r_0}\right)^3\right] \epsilon^2 \nonumber\\
	&\qquad + \mathcal{O}(\epsilon^3).\label{phi_pert2}
	\end{align}
\end{subequations}
Again the terms of the order $\epsilon^2$ in the metric can be considered as the background field. Note that with the choice of $C_1=0$, the leading order of $f(r)-1$, $h(r)-1$, and $\phi(r)$ in the above solution agrees with those in Ref.~\cite{Hossenfelder:2017eoh}. In particular, our agreement in Ref.\cite{Hossenfelder:2017eoh} at leading order occurs despite our differing equations of motion with \cite{Hossenfelder:2017eoh}, because the difference does not affect the leading order and \cite{Hossenfelder:2017eoh} did not consider back-reaction from the imposter field to the metric.

Thus, to obtain the pure gravitational effects caused by the impostor field, we should set $\delta =\epsilon^3$. To the subleading order, we get
\begin{subequations}
	\begin{align}
	f(r) &\sim 1 -\left[ C_1 + \frac{\alpha}{r} (2b-c) \sqrt{-(b+c)}  \ln\frac{r}{r_g} \right] (q\epsilon)^3 + \mathcal{O}(\epsilon^4), \label{f_pert3}\\
	h(r) &\sim 1 - \frac{\alpha}{r}\sqrt{-(b+c)} \left[-3b+(2b-c)\ln\frac{r}{r_g}\right] (q\epsilon)^3 +  \mathcal{O}(\epsilon^4), 
	\label{h_pert3}\\
	\phi(r) & \sim \left(\ln\frac{r}{r_0}\right) (q\epsilon) + \frac{2\lambda r^3}{243\alpha[-(b+c)]^{3/2}} \left[8 -18 \ln\frac{r}{r_0} +18 \left(\ln\frac{r}{r_0}\right)^2 -9 \left(\ln\frac{r}{r_0}\right)^3\right] (q\epsilon)^2 \nonumber\\
	&\qquad + \mathcal{O}(\epsilon^3),
	\label{phi_pert3}
	\end{align}
	\label{pert3}
\end{subequations} 
where we have replaced the integration constant $M$ by $r_g$. One surprising feature for this solution is that one of the integration constants, $q$ can be completely absorbed into the small parameter $\epsilon$. In the case without potential, $\Vcal=0$, we found that the metric $f(r)$ and $h(r)$ are power series of $\epsilon^3$ and the power expansion of the impostor field $\phi(r)$ has only terms in $\epsilon$, $\epsilon^4$, $\epsilon^7$, and so on [We actually computed all the way to the order of $\epsilon^{10}$, which are too long to be presented here.]. From the results, we observed that the solution for $f(r)$, $h(r)$, and $f(r)\phi(r)$ truncate for the choice $\lambda=C_1=b=0$. [And $f(r)=h(r)$ as a by-product.] This implies an exact solution for this choice of parameters. Indeed, such a solution exists and will be considered in further details in the following sections.

\subsection{Cosmological solution}

Next we apply the model to cosmology. 
We perform a perturbative analysis using a Robertson-Walker-type metric with a flat spatial part
\begin{subequations}
	\begin{align}
	\dif s^2&=-\dif t^2+\expo{2f(t)}\left[\dif r^2+r^2\dif\Omega^2_{(2)}\right],\\
	u^\mu&=\phi(t)\delta^\mu_t,
	\end{align}
\end{subequations}
with 
\begin{subequations}
	\begin{align}
	f(t) &\sim f_1(t) \delta+  \cdots, & \delta\to0,\\
	\phi(t) &\sim \phi_1(t) \epsilon + \cdots, &\epsilon\to0.
	\end{align}
\end{subequations}
With a wide range choices of the relative size of the two small parameters,  we found one and only consistent perturbative solution with $\delta=\epsilon^{3/2}$. This is in contrary to the case with a spherical symmetric metric. Up to $\mathcal{O}(\epsilon^4)$, we get 
\begin{subequations}
	\begin{align}
	f(t) &\sim  [2(a+b)]^{3/4} \sqrt{\frac{\alpha}{3}} t \epsilon^{3/2}  -\frac{3 b}{4}\sqrt{2(a+b)} \alpha t^2 \epsilon^3 -  \frac{8a-3b}{240(a+b)} \lambda t^6 \epsilon^4 + \mathcal{O}(\epsilon^{9/2}),\\
	\phi(t) & \sim t\epsilon - \frac{1}{30[2(a+b)]^{3/2}} \frac{\lambda t^5}{\alpha}  \epsilon^2- \frac{(2a+b)}{2[2(a+b)]^{1/4}} \sqrt{3\alpha} t^2 \epsilon^{5/2} -\frac{1}{12960(a+b)^3} \frac{\lambda^2t^9}{\alpha^2}  \epsilon^3 \nonumber\\
	  &\qquad + \frac{14a+3b}{60 [2(a+b)]^{7/4}} \frac{\lambda t^6}{\sqrt{3\alpha}}  \epsilon^{7/2} +\left[ \frac{8a^2+16ab+5b^2}{4\sqrt{2(a+b)}}\alpha t^3 - \frac{49\lambda^3t^{13}}{631800[2(a+b)]^{9/2}} \right] \epsilon^4 \nonumber\\
	  &\qquad + \mathcal{O}(\epsilon^{9/2}).
	\end{align}
\end{subequations}  
Unlike the spherical-symmetric solutions, there is only one integration constant in this solution. And this constant can be be absorbed into $\epsilon$ just like in Eq.~(\ref{pert3}). In this case, to ensure a real solution, we must have $a+b>0$. For this type of metric, the parameter $c$ disappears from the solution.

For all the perturbative solutions presented in this section, we found that stress-energy conservation is automatically satisfied. That is, other than $b+c<0$ and $a+b>0$, we do not have any other constraints on the parameters $a$, $b$, and $c$. The choice made in Refs.~\cite{Hossenfelder:2017eoh} and \cite{Dai:2017guq} is consistent with these constraints.

\section{Exact solution} 
\label{Ssolution}

The perturbative solutions of the previous section suggests that an exact solution to \Eqref{EOMu} and \Eqref{EOMg} contains an exact solution for the case $b=0$ and $c=-1$, where $T_{\mu\nu}=0$ and $\Vcal=0$. Indeed, one can explicitly verify that 
\begin{subequations}\label{solution}
\begin{align}
 \dif s^2&=-f(r)\dif t^2+\frac{\dif r^2}{f(r)}+r^2\dif\Omega^2_{(2)},\quad u^\mu=\phi(r)\,\delta^\mu_t,\label{soln_metric}\\
 f(r)&=1-\frac{2M}{r}-\frac{\alpha q^3}{r}\ln{\frac{r}{r_g}},\\
 \phi(r)&=\frac{q}{f(r)}\ln{\frac{r}{r_0}},
\end{align}
\end{subequations}
exactly solves \Eqref{EOMu} and \Eqref{EOMg}. This solution is parametrised by $M$, $q$, and $r_0$. The parameter $r_g$ can be absorbed by rescaling $M$, though we will keep it so that the argument of the logarithmic function appears explicitly dimensionless. 

Note that $a$ has been irrelevant in our spherically symmetric ansatz, since the kinetic term where $a$ is the coefficient identically vanishes. (Though it is non-trivial in the cosmological solutions.) However, the particular choice $a=b=0$ corresponds to an anti-symmetric combination for $\chi$, and thus the field is similar to that of a gauge boson. As such we might consider $\chi$ with $a=b=0$ to be something akin to a Maxwell-type entity; though the similarity ends if $a\neq0$. In fact, one may have already noticed that the metric \Eqref{soln_metric} is indeed similar to a black hole solution in non-linear Maxwell theory where the power of the Maxwell invariant is $3/2$ \cite{Maeda:2008ha, Gonzalez:2009nn}, which is precisely the power of $\chi$ in the Lagrangian. However, we emphasise that the $\chi$ is definitively not a Maxwell gauge field due to the presence of the term associated with $a$ which, in general, is non-zero.  

In any case, the imposter field $u$ should not be interpreted as an electromagnetic potential, as it couples to matter in a very different way --- clearly the field $u$ exerts forces on uncharged particles, as intended in the construction of emergent gravity. Unlike the vector potential in non-linear Maxwell theory, the imposter field $u$ will contribute to the gravitational potential that is argued to explain the galactic rotation curves in the dark matter problem \cite{Hossenfelder:2018vfs}. Furthermore, the inclusion of matter fields introduces $I_{\mathrm{int}}$ to the action, and will lead to very different results from non-linear Maxwell theory. We shall explore the effect of $u$ on particles in further detail in the following sections.

Before closing this section, let us briefly mention the physical properties of the spacetime. Firstly, the spacetime is asymptotically flat, though the metric functions include a term $\sim\ln(r)/r$ which dies off more slowly as compared to the pure Schwarzschild case. The horizon is located at $r=r_+$ for which $f(r_+)=0$. It may be more convenient to parametrise the solution with $r_+$ in place of $M$, where $M$ can be recovered by
\begin{align}
 M=\half\brac{r_+-\alpha q^3\ln\frac{r_+}{r_g}}.
\end{align}
In terms of $r_+$, the surface gravity and horizon area are given by 
\begin{align}
 \kappa=\frac{r_+-\alpha q^3}{2r_+^2},\quad \mathcal{A}=r_+^2\Omega_{(2)}, \label{kappa_and_A}
\end{align}
where $\Omega_{(2)}=4\pi$ is the area of a unit two-sphere. The Kretschmann invariant and Ricci scalar are respectively
\begin{align}
 R_{\rho\sigma\mu\nu}R^{\rho\sigma\mu\nu}&=\frac{1}{r^6}\Bigg\{  48M^2+8M\alpha q^3\brac{6\ln\frac{r}{r_g}-5}\nonumber\\
              &\quad\hspace{2cm}+\alpha^2q^6\sbrac{12\brac{\ln\frac{r}{r_g}}^2-20\ln\frac{r}{r_g}+13}\Bigg\},\\
 R&=\frac{\alpha q^3}{r^3},
\end{align}
indicating the presence of a curvature singularity at $r=0$. It can be easily checked that the solution \Eqref{solution} saturates the Null Energy Condition $R_{\mu\nu}k^\mu k^\nu\geq 0$ where $k^\mu$ is any null vector satisfying $k^\mu k_\mu=0$.

Since baryonic matter couples to the effective metric $\tilde{g}_{\mu\nu}$ as given by Eq.~\Eqref{tildeg}, it is also worth checking the behaviour of its curvature invariants as well. With the solution \Eqref{solution}, $\tilde{g}_{\mu\nu}$ is
\begin{align}
 \dif\tilde{s}^2&=-f\brac{1+\beta\sqrt{f}\phi}\dif t^2+f^{-1}\dif r^2+r^2\dif\Omega^2_{(2)}\nonumber\\
                &=-Nf\dif t^2+f^{-1}\dif r^2+r^2\dif\Omega^2_{(2)},
\end{align}
where we have denoted $N=\brac{1+\beta\sqrt{f}\phi}$. The metric appears to be singular at the roots $r_+$ and $r_*$, where $f(r_+)=0$ and $N(r_*)=0$. The Ricci scalar of $\tilde{g}_{\mu\nu}$ is
\begin{align}
 \tilde{R}&=-\frac{1}{r^2\sqrt{N}}\brac{r^2\frac{(Nf)'}{\sqrt{N}}}'-\frac{2}{r^2}\brac{f+rf'},
\end{align}
which diverges at $r=0$ and $r=r_*$. In particular, the latter case occurs when
\begin{align}
 \beta qf^{-1/2}\ln\brac{r/r_0}&=-1.
\end{align}
In the next section, we shall establish that a physically relevant range for $\beta$ and $q$ satisfies $\beta q>0$. Therefore $N=0$ may occur at some value of $r=r_*$ where $r_*/r_0<1$. Indeed, a curvature singularity $r=r_*$ seems unphysical if it occurs outside the horizon, and therefore we shall consider the cases where $r_*$ is hidden behind the horizon. In other words, for a given $M$, we require
\begin{align}
 r_*<r_+ \label{rstarrplus}
\end{align}
where particles experience a regular exterior metric. As such, Eq.~\Eqref{rstarrplus} 
gives an implicit constraint among the parameters $(\alpha,\beta,q,r_g,r_0)$ analogous to $a<M$ for Kerr black holes or $Q<M$ for Reissner-Nordstr\"{o}m black holes. A closer analogy would perhaps be the case of tensor-scalar-vector gravity, where an effective metric was also used and the parameters are constrained so as to avoid curvature singularities \cite{Giannios:2005es}.

\section{Motion of test masses} \label{Stestmass}

\subsection{Equations of motion and the Newtonian limit}

As discussed in Sec.~\ref{Saction}, Hossenfelder's model \cite{Hossenfelder:2017eoh} was designed to be a covariant description of \cite{Verlinde:2016toy}, where matter feels an effective metric of the form $\tilde{g}_{\mu\nu}=g_{\mu\nu}-\beta\frac{u_\mu u_\nu}{u}$. Therefore, the motion of a test particle is no longer a geodesic curve of $g_{\mu\nu}$, but rather that of the \emph{effective} metric $\tilde{g}_{\mu\nu}$. This is among the main difference between CEG and non-linear Maxwell theory, even though both contain exact solutions given by the metric in \Eqref{solution}. In particular, the imposter field exerts a force directly on uncharged particles via the interaction $\mathcal{L}_{\mathrm{int}}$, which is why particles experience an effective metric $\tilde{g}_{\mu\nu}$ instead of moving in `force-free' geodesics in $g_{\mu\nu}$. 

In light of this, for the motion of particles described by a curve $x^\mu(\tau)$ parametrised by $\tau$. As we have mentioned in Sec.~\ref{Saction}, the action for a particle of mass $m$ is described by Eq.~\Eqref{TestParticleAction}. Let us write the Lagrangian here for convenience:
\begin{align}
 \mathcal{L}&=\frac{m}{2}\tilde{g}_{\mu\nu}\dot{x}^\mu\dot{x}^\nu=\frac{m}{2}\brac{g_{\mu\nu}\dot{x}^\mu\dot{x}^\nu-\beta\frac{u_\mu u_\nu}{u}\dot{x}^\mu\dot{x}^\nu},
\end{align}
where over-dots denote derivatives with respect to $\tau$. Applying the Euler-Lagrange equation, or equivalently, extremising $\int\dif\tau\mathcal{L}$, leads to
\begin{align}
 \brac{\delta^\kappa_\mu-\beta\frac{u_\mu u^\kappa}{u}}\ddot{x}^\mu+\Gamma^\kappa_{\mu\nu}\dot{x}^\mu\dot{x}^\nu=\beta{C^\kappa}_{\mu\nu}\dot{x}^\mu\dot{x}^\nu, \label{GeodesicEqn}
\end{align}
where 
\begin{align}
 {C^{\kappa}}_{\mu\nu}=\half g^{\kappa\lambda}\sbrac{\partial_\mu\brac{\frac{u_\lambda u_\nu}{u}}+\partial_\nu\brac{\frac{u_\lambda u_\mu}{u}}-\partial_\lambda\brac{\frac{u_\mu u_\nu}{u}}}
\end{align}

At this stage, it is important to reiterate that $x^\mu(\tau)$ that solves Eq.~\Eqref{GeodesicEqn} is a geodesic of an \emph{effective} metric $\tilde{g}_{\mu\nu}=g_{\mu\nu}-\beta\frac{u_\mu u_\nu}{u}$, and therefore $\dot{x}^\mu$ is no longer a parallel-transported vector in the spacetime $g_{\mu\nu}$. Nevertheless, as a geodesic of $\tilde{g}_{\mu\nu}$, it is a vector that is parallel-transported along that effective metric, and therefore the inner product $\tilde{g}_{\mu\nu}\dot{x}^\mu\dot{x}^\nu$ is constant along the geodesic. For time-like geodesics, we can then always rescale $\tau$ to make the magnitude of this constant be unity, giving us a first integral
\begin{align}
 \brac{g_{\mu\nu}-\beta\frac{u_\mu u_\nu}{u}}\dot{x}^\mu\dot{x}^\nu=-1. \label{timelike_epsilon}
\end{align}

We can take the non-relativistic limit by having $\dot{t}\gg\dot{x}^i$. Further assuming the spacetime is static and the $tt$-component of the metric is of the form $g_{tt}\simeq -1+2\psi_{\mathrm{N}}$, Eq.~\Eqref{GeodesicEqn} reduces to 
\begin{align}
 m\frac{\dif^2\vec{r}}{\dif t^2}&=-\vec{\nabla}\brac{\psi_{\mathrm{N}}+\psi_u}.
\end{align}
where $\psi_u$ is the contribution from the imposter field. Taking the lowest orders of the functions in \Eqref{solution}, the potentials are
\begin{align}
 \psi_{\mathrm{N}}&=\frac{M}{r}+\half\alpha q^3\frac{\ln\brac{r/r_g}}{r},\\
 \psi_u&=-\frac{\beta q}{2}\ln\frac{r}{r_0}.
\end{align}
We see that $\psi_{\mathrm{N}}$ is the gravitational potential obtained by taking the weak field limit of the spacetime geometry, and $\psi_u$ is the potential due to the imposter field directly exerting a force on the moving particles. Interestingly, since this limit is obtained from the exact solution where the mass-energy of the imposter field contributes to the spacetime curvature, $\psi_{\mathrm{N}}$ has an additional term $\half\alpha q^3\frac{\ln\brac{r/r_g}}{r}$, which is a relativistic correction coming from this back-reaction. Incidentally, a $\ln(r)/r$ term was also obtained by \cite{Cadoni:2017evg} is reproduced within $\psi_{\mathrm{N}}$ from a different context.

Differentiating the total potential $\psi=\psi_{\mathrm{N}}+\psi_u$, we find that the acceleration of the particle is
\begin{align}
 a=-\frac{M}{r^2}-\frac{\beta q}{2r}+\frac{\alpha q^3}{2r^2}-\frac{\alpha q^3}{2r^2}\ln\brac{\frac{r}{r_g}}.\label{MOND}
\end{align}
Although we have derived this result starting from the $b=0$ exact solution \Eqref{solution}, Eq.~\Eqref{MOND} can be modified easily for generic $a$, $b$, and $c$ based on the perturbative solutions in Sec.~\ref{SSssph}.

Looking at Eq.~\Eqref{MOND} more closely, we see that the first two terms reproduces the MOND relation obtained in \cite{Hossenfelder:2018vfs} with the MOND parameter expressed in terms of the present notation as
\begin{align}
 \sqrt{a_0}=\frac{\beta q}{2\sqrt{M}}, \label{MOND_parameter}
\end{align}
and the last two terms are due to the backreaction of the imposter field to the metric, which will be negligible if $\alpha q^3$ is sufficiently small. Let us then regard these terms as the relativistic correction to MOND under CEG.

Having this non-relativistic limit allows us to make a few statements on some of the parameters of our solution. Firstly, for Eq.~\Eqref{MOND} to appropriately contribute to the galactic rotation curves, we require $\beta q>0$. To further make contact with the Lagrangian in \cite{Hossenfelder:2017eoh}, we have $\beta=1/L$. This sets $q$ to be positive. The quantity $\alpha q^3$ may perhaps be constrained by the rotation curve data of the galaxies via Eq.~\Eqref{MOND}.  However, if we were to follow \cite{Hossenfelder:2017eoh}, the coefficient of $\chi^{3/2}$ in the Lagrangian is positive, and we shall henceforth consider $\alpha>0$.

\subsection{Relativistic test mass}

We now consider the test mass in a fully relativistic treatment. For the exact solution \Eqref{solution}, the particle Lagrangian is 
\begin{align}
 \mathcal{L}&=\frac{m}{2}\sbrac{-\brac{f+\beta f^{3/2}\phi }\dot{t}^2+\frac{\dot{r}^2}{f}+r^2\dot{\theta}^2+r^2\sin^2\theta\,\dot{\varphi}^2}.\label{timelike_Lagrangian}
\end{align}
We can use the spherical symmetry of the spacetime to fix the coordinate system such that the motion is confined to the plane $\theta=\frac{\pi}{2}=\mathrm{constant}$, and we need not consider $\theta$ henceforth. 

With the Lagrangian \Eqref{timelike_Lagrangian}, we can now proceed to obtain the equations of motion. Since $t$ and $\varphi$ are cyclic variables, we have the first integrals
\begin{align}
 \dot{t}=\frac{E}{f\brac{1+\beta\sqrt{f}\phi}},\quad\dot{\varphi}=\frac{L}{r^2},
\end{align}
where $E$ and $L$ may be interpreted as the energy and angular momentum of the particle, respectively. Applying the Euler-Lagrange equation to the coordinate $r$ gives 
\begin{align}
 \ddot{r}&=\frac{f'}{2f}\dot{r}^2-\frac{\brac{f+\beta f^{3/2}\phi}'}{2f\brac{1+\beta\sqrt{f}\phi}^2}E^2+\frac{L^2f}{r^3}.
\end{align}
As argued in the previous subsection, inner products of vectors are preserved if they are parallel-transported in $\tilde{g}_{\mu\nu}$ instead of $g_{\mu\nu}$. Therefore Eq.~\Eqref{timelike_epsilon} gives a constraint
\begin{align}
 \dot{r}^2&=\frac{E^2}{1+\beta \sqrt{f}\phi}-\brac{\frac{L^2}{r^2}+1}f. \label{timelike_first_integral}
\end{align}

Because of the logarithmic functions appearing in $f$ and $\phi$, it will generally be difficult to integrate the equations of motion exactly. Nevertheless we could still extract some qualitative features by inspecting Eq.~\Eqref{timelike_first_integral}, in addition to solving the equations numerically.

\begin{figure}
 \begin{center}
 \includegraphics{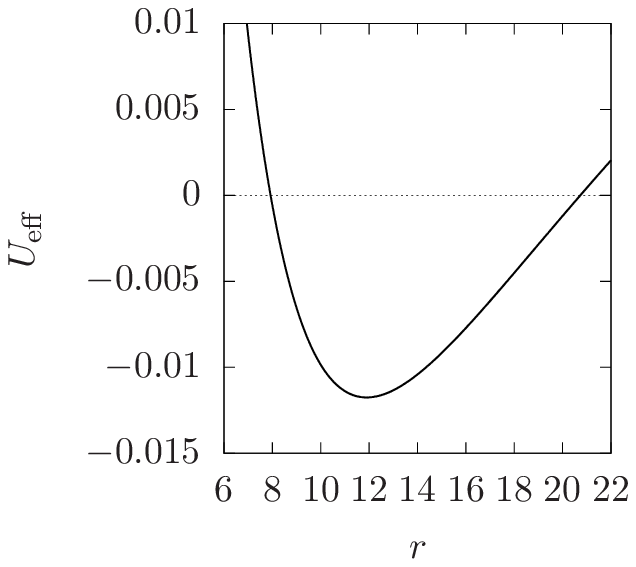}
 \includegraphics{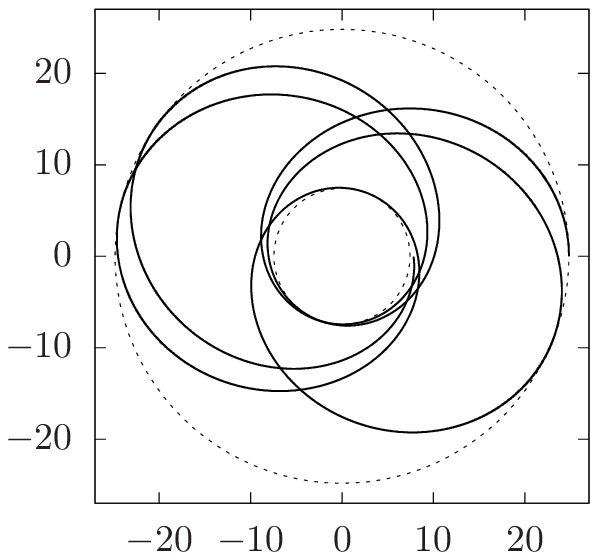}
 \caption{A test mass trajectory of energy $E^2=0.94$ and $L^2=16$ in a spacetime of parameters $M=1$, $r_0=1$, $q=1$, $\alpha=0.001$, and $\beta=0.001$. The left plot shows $U_{\mathrm{eff}}  $ vs $r$, where negative $U_{\mathrm{eff}}  $ is seen to lie in the range $7.44<r<24.8$. And this range can be clearly seen in the trajectory plotted on the right in Cartesian coordinates.}
 \label{fig_01rdot_range_example}
 \end{center}
\end{figure}

For instance, given a particle of a specific energy and angular momentum in a spacetime with parameters $M$, $q$, $\alpha$, $\beta$, and $r_0$, we can look for the presence of stable bound orbits by finding a finite range where $\dot{r}^2>0$ in Eq.~\Eqref{timelike_first_integral}. We can equivalently express this in terms of an effective potential
\begin{align}
 U_{\mathrm{eff}}\equiv-\dot{r}^2=-\frac{E^2}{1+\beta \sqrt{f}\phi}+\brac{\frac{L^2}{r^2}+1}f,
\end{align}
in which allowed orbits are expressed as $U_{\mathrm{eff}}<0$. If this range includes the horizon $r_+$, then this particle may eventually fall into the black hole. On the other hand, if the range extends to infinity, the particle is unbound and may escape. However if we find a finite range of $r$ with negative $U_{\mathrm{eff}}$, the particle is in a stable bound orbit. An example of a bound orbit is shown in Fig.~\ref{fig_01rdot_range_example}.

\begin{figure}
 \begin{center}
 \begin{subfigure}[b]{0.49\textwidth}
   \centering
   \includegraphics[scale=0.86]{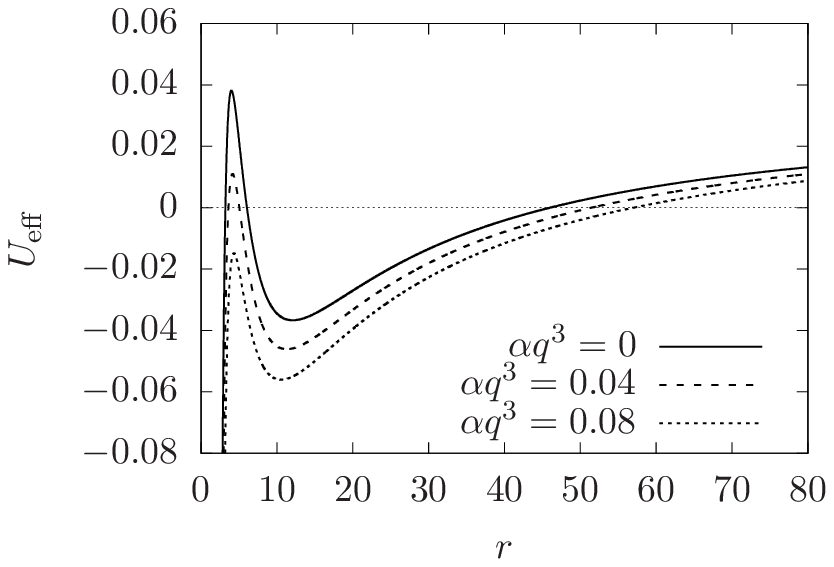}
   \caption{$\beta q=0.001$.}
   \label{fig_02rdot_bound_orbits_vary_alpha}
 \end{subfigure}
 \begin{subfigure}[b]{0.49\textwidth}
   \centering
   \includegraphics[scale=0.86]{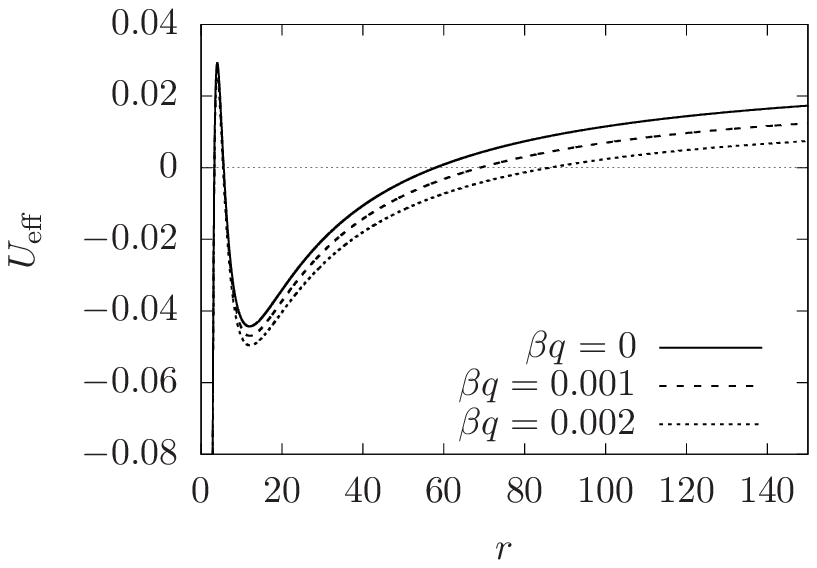}
   \caption{$\alpha q^3=0.001$.}
   \label{fig_02rdot_bound_orbits_vary_beta}
 \end{subfigure} 
 \caption{Plots of $U_{\mathrm{eff}}  $ vs $r$ for trajectories with angular momentum $L^2=16$ and energy $E^2=0.97$ in a spacetime of parameters $M=r_g=r_0=1$. Fig.~\ref{fig_02rdot_bound_orbits_vary_alpha} shows the effective potential for fixed $\beta q$, whereas Fig.~\ref{fig_02rdot_bound_orbits_vary_beta} is for fixed $\alpha q^3$.}
 \label{fig_02rdot_bound_orbits}
 \end{center}
\end{figure}

With the effective potential at hand, we can explore some qualitative features of bound orbits in the spacetime. A numerical exploration of the parameters shows that increasing $\alpha q^3$ tends to lower the potential barrier from a similar Schwarzschild orbit with the same energy and angular momentum. Some representative shapes of the effective potentials are shown in Fig.~\ref{fig_02rdot_bound_orbits}. In particular, we see in Fig.~\ref{fig_02rdot_bound_orbits_vary_alpha} that the potential barrier is also lowered where for sufficiently large $\alpha q^3$, the peak of the potential dips below $U_{\mathrm{eff}}  =0$, and the particle may fall into the horizon. At the same time, increasing $\alpha q^3$ also widens the range $r$ where $U_{\mathrm{eff}} <0$. 

On the other hand, we see that increasing $\beta q$ from zero shows a more pronounced widening of the $U_{\mathrm{eff}}  <0$ range, whilst the potential barrier roughly fixed for fixed $\alpha$, as shown in Fig.~\ref{fig_02rdot_bound_orbits_vary_beta}. Therefore, for CEG-related parameters with positive $\alpha q^3$ and $\beta q$, particles with the same energy and angular momentum have orbits that reach further out from the black hole as compared to the Schwarzschild case.

\begin{figure}
 \begin{center}
  \includegraphics{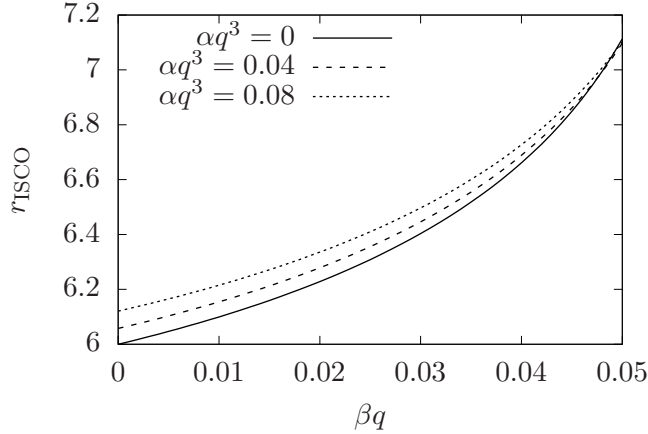}
  \caption{Plots of the innermost stable circular orbits, $r_{\mathrm{ISCO}}$ against $\beta$ for various $\alpha=0$, $0.04$, and $0.08$, for a spacetime with $M=r_g=r_0=1$.}
  \label{fig_02bISCO}
 \end{center}
\end{figure}

A property of particle orbits which potentially has some observational relevance is perhaps the \emph{innermost stable circular orbits} (ISCOs). The size of the ISCO may affect the bounds of spectral line broadening of radiation emitted by charged particles orbiting a magnetised central body \cite{Frolov:2014zia,Kolos:2017ojf}.
An ISCO is an orbit where the unstable maxima and stable minima of the potential coincide, satisfying 
\begin{align}
 \frac{\dif U_{\mathrm{eff}}  }{\dif r}=0,\quad \frac{\dif^2 U_{\mathrm{eff}}  }{\dif r^2}=0.\label{ISCO_eqn} 
\end{align}
To find an ISCO, we simply solve the simultaneous equations \Eqref{ISCO_eqn} for $r$ and $L$. Let us denote the solution for $r$ as $r_{\mathrm{ISCO}}$. While this quantity is too cumbersome to be presented in closed-form, we can easily see how its value changes for different $\alpha q^3$ and $\beta q$. Again, by numerical exploration, we find that increasing both parameters tend to increase $r_{\mathrm{ISCO}}$. Some representative examples are shown in Fig.~\ref{fig_02bISCO}.\footnote{Recall that in the Schwarzschild case, $r_{\mathrm{ISCO}}=6M$.}

\section{Motion of photons}\label{Sphotons}

At this stage, Refs.~\cite{Hossenfelder:2017eoh,Verlinde:2016toy} do not indicate how massless fields couple to $u^\mu$. Lacking further information, it seems reasonable to assume that photons still travel along geodesics of the physical spacetime, and we have 
\begin{align}
 g_{\mu\nu}\dot{x}^\mu\dot{x}^\nu=0.\label{null_epsilon}
\end{align}

The corresponding Lagrangian is simply $2\mathcal{L}=g_{\mu\nu}\dot{x}^\mu\dot{x}^\nu$ per unit energy. For the metric \Eqref{soln_metric}
\begin{align}
 \mathcal{L}&=\half\brac{-f\dot{t}^2+\frac{\dot{r}^2}{f}+r^2\dot{\varphi}^2},
\end{align}
where again, we use the spherical symmetry of the spacetime to fix the coordinates such that the geodesics is confined to the plane $\theta=\frac{\pi}{2}=\mbox{constant}$. As in the previous subsection, we can derive the equations of motion in a similar manner, which gives us
\begin{subequations}\label{photon_eom}
\begin{align}
 \dot{t}&=\frac{E}{f},\quad\dot{\varphi}=\frac{L}{r^2},\\
 \dot{r}^2&=E^2-\frac{L^2}{r^2}f,\\
 \ddot{r}&=\frac{f'}{2f^2}\dot{r}^2-\frac{f'E^2}{2f}+\frac{fL^2}{r^2}.
\end{align}
\end{subequations}

\begin{figure}
 \begin{center}
  \includegraphics{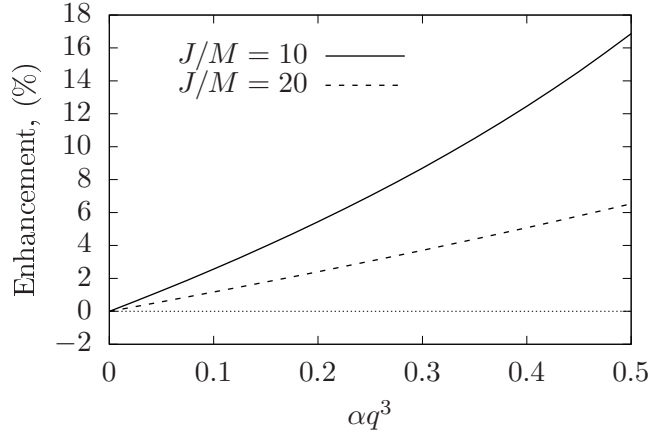}
  \caption{Percentage of bending-angle enhancement over standard GR, for impact parameters $J/M=10$ and $J/M=20$.}
  \label{fig_03lensing}
 \end{center}
\end{figure}

We can find the (coordinate) distance of closest approach, $r_{\mathrm{min}}$ by finding the largest root of $r$ satisfying $\dot{r}=0$. This gives 
\begin{align}
 \frac{E^2}{L^2}=\frac{f(r_{\mathrm{min}})}{r_{\mathrm{min}}^2}.
\end{align}
Further defining $u=1/r$ and $u_{\mathrm{min}}=1/r_{\mathrm{min}}$, we can calculate the deflection angle by dividing $\dot{r}$ with $\dot{\varphi}$ from Eq.~\Eqref{photon_eom} and integrating
\begin{align}
 \Delta\varphi&=2\int_{0}^{u_{\mathrm{min}}}\frac{\dif u}{\sqrt{u_{\mathrm{min}}^2f(1/u_{\mathrm{min}})-u^2f(1/u)}}.
\end{align}
In order to compare how lensing in CEG fares against the Schwarzschild ($\alpha q^3=0$) case, we calculate lensing in both cases for photons with the same impact parameter 
\begin{align}
 J=\frac{r_{\mathrm{min}}}{\sqrt{f(r_{\mathrm{min}})}}.
\end{align}
The numerical results can be seen in Fig.~\ref{fig_03lensing}, where as $\alpha q^3$ is increased, lensing is greater than the Schwarzschild case at $\alpha q^3=0$. As expected, the smaller impact parameter results in a larger bending angle, as the photon passes within a closer proximity to the gravitating mass. 

For fixed $M$, the effects of $\alpha q^3$ are more pronounced for smaller impact parameter, i.e., the enhancement of $\Delta\varphi$ by $\alpha q^3$ is greater for smaller $J/M$. For instance, in Fig.~\ref{fig_03lensing}, we depict the enhancement of the bending angle over standard GR, calculated using $(\Delta\varphi-\Delta\varphi_{\mathrm{Sch}})/\Delta\varphi_{\mathrm{Sch}}\times 100\%$, at $J/M=10$, increasing $\alpha q^3$ from zero to 0.5 increases the bending angle by roughly $0.6$ radians from the Schwarzschild value. But if the impact parameter is increased to $J/M=20$, the same range of $\alpha q^3$ only enhances the angle by around $0.2$ radians. If $J/M$ is too large, it may be too insensitive to $\alpha q^3$ and the lensing may be indistinguishable from standard GR. 

We can also further consider what are the possible observational consequences of the above calculations. As mentioned in the previous paragraph, the bending angle is more sensitive to the $\alpha q^3$ for smaller $J/M$. The latter depends on the particular lensing scenario, i.e., whether the observed lensing is due to a stellar-mass object or strong lensing by a galactic cluster. To get a rough estimation on the physically relevant size of $J/M$, we take, for instance, data from galaxy cluster lensing \cite{Allen:1997vc}, and lensing by individual clusters \cite{Bolton:2005nf,Horvath:2011xr}. 

\begin{figure}
 \begin{center}
  \includegraphics{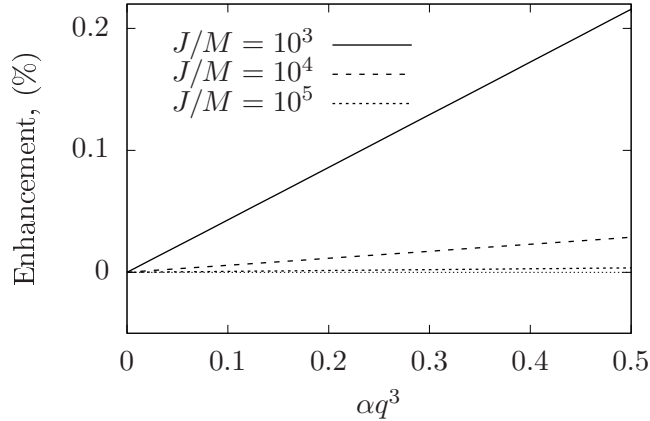}
  \caption{Percentage of bending-angle enhancement over standard GR for against $\alpha q^3$, for $J/M=10^3$, $10^4$, and $10^5$.}
  \label{fig_04clusters}
 \end{center}

\end{figure}

In any case, we shall estimate $J/M$ for clusters by assuming $r_{\mathrm{arc}}\simeq J$, where $r_{\mathrm{arc}}$ is the arc radius of the lens image provided in \cite{Allen:1997vc}. We take the mass of the cluster to be $M_X$ in \cite{Allen:1997vc}, which is measured from its X-ray emissions. In Ref.~\cite{Allen:1997vc}, $M_X$ does not agree with $M_{\mathrm{arc}}$ where the latter is determined from the lensing model under standard GR. For the case of individual galaxy lensing, we select a few samples from the Sloan Digital Sky Survey \cite{Bolton:2005nf}, using the notation of \cite{Horvath:2011xr}. For this case, we take $r_{\mathrm{min}}\simeq J$, and their masses are given by \cite{Grillo}. The values of $J/M$ are shown in Table \ref{tab_JM}, where we see a typical $J/M$ ratio is around $10^3\sim10^5$. Having this range of $J/M$ in mind, we can narrow down the amount of lensing that is enhanced over that of standard GR. We show the cases $J/M=10^3$, $J/M=10^4$, and $J/M=10^5$ in Fig.~\ref{fig_04clusters}. More precisely, the amount of enhancement over standard GR is roughly around $10^{-5}\sim 10^{-3}$ radians, which correspond to the order of $1\sim 100$ arc-seconds.

\begin{table}
 \begin{center}
 \begin{tabular}{llllllc}
  Object             & $J$ (kpc) & mass ($M_\odot)$  & $J/M$ &Reference\\[6pt]\hline\\[-9pt]
  PKS0745-191        & 45.9 & $3.16\times 10^{13}$ &   $\sim3.0\times 10^{4}$  & \cite{Allen:1997vc}\\[3pt]
  RXJ1347.5-1145     & 240 & $68.1\times 10^{13}$ &   $\sim7.4\times 10^{3}$  & \cite{Allen:1997vc}\\[3pt]
  MS1358.4+6245      & 121 & $7.03\times 10^{13}$ &  $\sim4.7\times 10^{3}$  & \cite{Allen:1997vc}\\[3pt]
  Abell 2744         & 119.6 & $2.78\times 10^{13}$  & $\sim9.0\times 10^{4}$ & \cite{Allen:1997vc}\\[3pt]
  Abell 2163         & 67.7 & $1.74\times 10^{13}$ & $\sim8.1\times 10^{5}$ & \cite{Allen:1997vc}\\[3pt]
  Abell 2218         & 79.4 & $1.89\times 10^{13}$ &  $\sim8.8\times 10^{4}$ & \cite{Allen:1997vc}\\[3pt]
  J0008 - 0004       & 6.595 & $1.30\times 10^{10}$ & $\sim3.9\times 10^{5}$ & \cite{Horvath:2011xr}\\[3pt]
  J0903 - 4116       & 7.237 & $45.0\times 10^{10}$ & $\sim3.4\times 10^{5}$ & \cite{Horvath:2011xr}\\[3pt]
  \hline
 \end{tabular}
 \caption{Values of $J/M$ for some gravitational lens samples. The lenses from Ref.~\cite{Allen:1997vc} are galaxy clusters, which can be seen to be typically $10^3$ times more massive than the single-galaxy lenses analysed in Ref.~\cite{Horvath:2011xr}. To obtain the dimensionless ratio $J/M$, the galaxy masses in units of solar mass are to be converted to $M$ in geometrical units.}
 \label{tab_JM}
 \end{center}
\end{table}

\section{Conclusion} \label{Sconclusion}

With a fully relativistic variation of the action, we derive a full set of field equations from the CEG Lagrangian. We find that our resulting equations of motion involves a stress tensor that is different from \cite{Hossenfelder:2017eoh} and \cite{Dai:2017guq}, where our present stress tensor has an additional contribution due to the variation $\delta\Gamma^\lambda_{\mu\nu}$.

In the case where $b=0$, we considered an exact solution corresponding to a static, spherically-symmetric spacetime and an imposter field that varies logarithmically in $r$. While this system corresponds to a different stress tensor considered by \cite{Hossenfelder:2017eoh} and \cite{Dai:2017guq}, the weak-field limit nevertheless reproduces the MOND relations in \cite{Hossenfelder:2017eoh,Hossenfelder:2018vfs}, along with additional relativistic corrections which was not captured in its purely non-relativistic derivation. Furthermore, the perturbative solutions in Sec.~\ref{SSssph} also reproduces the MOND potential. We have also calculated the motion of fully relativistic test masses and gravitational lensing.

Verlinde's main formulation for emergent gravity comes from arguments of entropy. In the present paper, we have obtained a black hole solution with its associated surface gravity and horizon area given in Eq.~\Eqref{kappa_and_A}. It would be interesting to consider a thermodynamic analysis of this solution, perhaps along the lines of the Gibbons-Hawking path-integral method \cite{Gibbons:1976ue}. Indeed, since the metric \Eqref{soln_metric} is similar to a black hole in non-linear Maxwell theory, some of the thermodynamic analysis of Gonzales et al. \cite{Gonzalez:2009nn} could be carried over. 
Furthermore, a non-trivial $\Vcal(u)$ should probably be taken into account as well.

With regards to gravitational lensing, we have so far assumed that photons travel along null geodesics of the metric in the usual manner. In this view, the imposter field do not directly exert forces on photons, but only influence their motion indirectly through its backreaction on the metric. Our solution is asymptotically flat with no cosmological horizons. Thus we take the coordinate angle $\Delta\varphi$ to be equivalent to the angle measured by an observer at infinity, therefore we need not measure bending angles at finite distances, thus avoiding the need to apply the Rindler-Ishak method \cite{Rindler:2007zz}.

While we have demonstrated that the presence of an imposter field provides a positive contribution to the bending angle, it should be noted that this is in the somewhat idealised case of a static, spherically-symmetric vacuum solution. Lensing observations are due to galaxy or galaxy clusters with non-trivial mass distribution. Furthermore, in order to draw conclusions of the theory in relation to observation, the redshift has to be taken into account. 

On the other hand, one could avoid the cosmological effects by considering the possibility of observing lensing by a compact object within our galaxy. The analysis in Sec.~\ref{Sphotons} can be extended to calculate the predicted results of a possible direct observation via long base-line interferometry \cite{Hirabayashi:2005kc,Doeleman:2008qh}, which is hoped to be possible within the near future \cite{Castelvecchi:2017}. However, qualitatively speaking, the CEG model has been shown to fit the galaxy rotation curve along the lines of MOND. Therefore it remains to be seen whether we expect any significant difference from standard GR in the sub-galactic scale.

\appendix

\bibliographystyle{ceg}

\bibliography{ceg}

\end{document}